\documentstyle[12pt,epsfig,cite]{article}

\newlength{\dinwidth}
\newlength{\dinmargin}
\def\docnum#1{\hbox to \hsize{\hskip123mm\hbox{#1}\hss}}
\def\date#1{\edef\@temp{#1}\ifx\@temp\@empty\def\@temp{\today}\fi
\hbox to \hsize{\hskip123mm\hbox{\@temp}\hss}}
\def\title#1{\vskip 0.8in plus 2in\begin{center}%
{\Large\bf#1\par}\vskip1.5em\end{center}\vskip 1in}
\def\@makefnmark{\hbox{$^{\@thefnmark)}$}}
\def\author#1{
\setcounter{footnote}{0}\def\@currentlabel{}%
\begingroup\def\thefootnote{\arabic{footnote}}
\def\@makefnmark{\hbox{$^{\@thefnmark)}$}}
\global\@topnum\z@ \large\begin{center}{\lineskip.5em
\begin{tabular}[t]{c}#1\end{tabular}\par}
\end{center}\par\vskip1.5em\@thanks\endgroup}

\def\abstract{\vskip0.8in plus 3in\begin{center}{\large\bf Abstract}\end{center}\quotation}

\def\d{{\rm d}}

\def\e{{\rm e}}
\def\i{{\rm i}}

\def\ee{e$^+$e$^-\;$}

\def\uu{${\rm u}\bar{\rm u}\;$}
\def\dd{${\rm d}\bar{\rm d}\;$}

\begin{document}
\begin{titlepage}
\title 
      {Strangeness production in a statistical
effective model of hadronisation}

\centerline{\large F. Becattini*, G. Pettini}
\vspace{0.5cm} 
\centerline{\it University of Florence and INFN Sezione di Firenze} 
\centerline{\it Largo E. Fermi 2, I-50125}
\centerline{\it Firenze (Italy)}
\centerline{becattini@fi.infn.it, pettini@fi.infn.it}

\begin{abstract}
We suppose that overall strangeness production in both high energy 
elementary and heavy ion collisions can be described within the framework 
of an equilibrium statistical model in which the effective degrees of freedom 
are constituent quarks as used in effective lagrangian models. In 
this picture, the excess of relative strangeness production in heavy ion 
collisions with respect to elementary particle collisions arises from the 
unbalance between initial non-strange matter and antimatter and from the
exact colour and flavour quantum number conservation over different finite
volumes. The comparison with the data and the possible sources of model 
dependence are discussed. 
\end{abstract}

\vspace*{1.5cm}
\centerline{\it To be published in the Proceedings of the conference}
\centerline{\it QCD@work, Martina Franca (Italy), June 16-20 2001}

\end{titlepage}

\section{Introduction: strangeness production and statistical hadronisation}

Recent observations have shown that hadron multiplicities in both \ee
and hadronic high energy collisions agree very well with a statistical-thermal 
ansatz \cite{beca}. This finding has been interpreted in terms of a multi-cluster 
hadronisation process in which each cluster fills its relevant multi-hadronic 
phase space in a pure statistical fashion, at a critical value of energy density 
\cite{beca, beca2}. 
Within this framework, temperature and other thermodynamical quantities have 
an essential statistical meaning which does not imply the existence of a 
thermalisation process at hadronic level through multiple collisions; rather,
hadronisation itself yields a statistically equilibrated hadronic population. 
One of the main features of this approach is the very low number of free 
parameters required. Under suitable assumptions about cluster masses and charges 
fluctuations at fixed volumes \cite{beca, beca2}, there are essentially two 
free parameters, namely the sum $V$ of the volumes of the clusters, and the 
temperature $T$. Yet, in order to reproduce the yields of strange particles, 
the model has to be supplemented with one more phenomenological parameter, 
$\gamma_S$, which suppresses the production of particles containing $n$ strange 
quarks by a factor $\gamma_S^n$ \footnote{very recently \cite{beca2} a new 
parametrisation of extra strangeness suppression has been proposed which is 
equivalent to that with $\gamma_S$ at very high multiplicity}. From fits to 
the available data, $\gamma_S$ turns out to be $<1$ in all examined collisions 
and strongly dependent on the initial colliding systems 
\cite{beca, beca2, becahion} so that full strangeness chemical equilibrium
is never observed.

A further insight in strangeness production is achieved by calculating the 
ratio between newly produced valence strange quarks and u, d quarks, the so-called
Wroblewski factor $\lambda_S = 2 \langle {\rm s \bar s} \rangle / (\langle 
{\rm u \bar u} \rangle + \langle {\rm d \bar d} \rangle)$, which is in fact 
fairly constant in all kinds of elementary collisions (EC) over a centre-of-mass 
energy range spanning about two orders of magnitude, while it shows a non-trivial 
behaviour in heavy ion collisions (HIC) and it is as twice as large at 
$\sqrt s \approx 20$ GeV \cite{becahion} (see Fig.~1). It should be mentioned 
that this ratio is calculated by using the fitted primary (i.e. directly emitted 
from the hadronising source) hadron multiplicities before hadronic decays take place,
which are not measurable. Thus, this parameter depend in fact on the 
fitted parameters $T$, $V$ and $\gamma_S$ and, in principle, is a model dependent 
quantity; still, the model accurately reproduces all measured multiplicities 
and the model dependence decreases as the number of measured final multiplicities 
increases, so that the estimated $\lambda_S$ is expected to be reasonably close 
to its actual value. 
\begin{figure}
\begin{center}
  \resizebox{20pc}{!}{\includegraphics{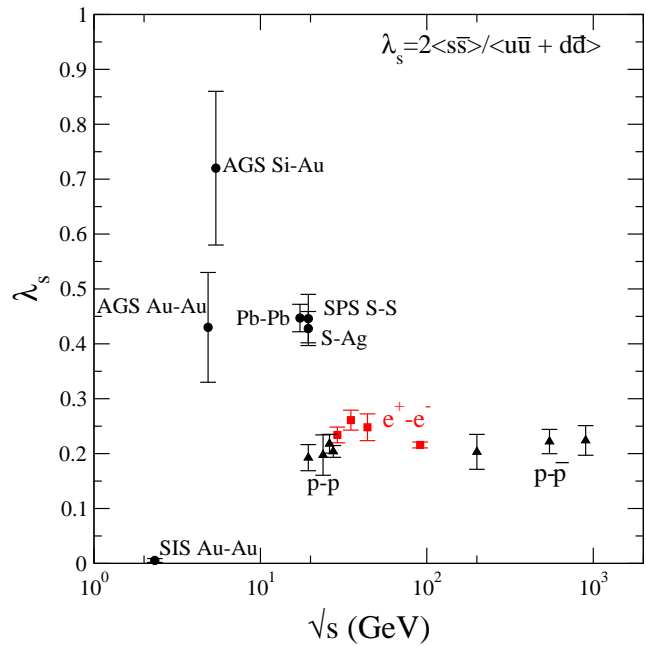}}
  \caption{$\lambda_S$ as a function of centre-of-mass energy in several kinds
  of collisions (from ref.~\cite{becahion})}
\end{center}  
\end{figure}
The explanation of the strangeness production in high energy collisions, and
especially $\lambda_S$, is a major goal for the understanding of hadronisation
and of the possible QGP formation in HIC.
 
In the following, we will try to account for this behaviour by resorting to a
statistical description in terms of consituent quarks rather than hadrons, or, 
more specifically, using microscopic models containing quarks as fundamental 
degrees of freedom. As full QCD cannot be handled, we use effective models (EM)
at finite temperature, which are commonly employed to investigate QCD phase 
transition at high temperatures and low baryon chemical potential, that is
our region of interest. Particularly, we will refer to low energy models with 
four-fermion interactions such as the Nambu-Jona-Lasinio model \cite{njlor,njl} 
and to what is sometimes referred to, in literature, as ${\it ladder}$-QCD 
\cite{bcd}.

\section{The model}

In using effective models with quarks as fundamental degrees of freedom to
calculate relative flavour production, essentially the same physical scheme of 
the statistical hadronisation model (SHM) both in HIC and EC is kept. This means 
that the formation of a set of hadronising clusters endowed with mass, volume 
and flavour quantum charges is assumed, in which every allowed quantum state 
is equally likely. Also the assumption of suitable maximum-disorder fluctuations 
of cluster flavour charges and masses for fixed volumes, enabling the introduction 
of a global temperature \cite{beca2}, is retained. Three further
assumptions are introduced:      

\begin{enumerate}
\item{}Each single cluster is a colour singlet
\item{}The produced s quarks, or at least the ratio s/u, survive in the 
hadronic phase
\item{}The temperature $T$ and the chemical potentials (in the grand-canonical 
framework) fitted in the SHM with hadron multiplicities are interpreted as 
the critical values for deconfinement and (approximate) chiral symmetry restoration
\end{enumerate}

The physical process of single cluster hadronisation, in both EC and HIC is 
envisaged as an evolution towards a chaotic quantum state which finally leads to 
statistical multi-hadronic phase space population by coalescence of produced 
constituent quarks. Within this picture, the lack of complete strangeness chemical
equilibrium at hadron level is the effect of a complete strangeness chemical 
equilibrium at constituent quark level. Whilst in HIC an {\em early} thermalisation 
and colour deconfinement over a large volume is expected, with consequent formation 
of relatively {\em large} colour singlet clusters, in EC the chaotic behaviour of 
quantum dynamics is supposed to set in at a 
{\em late} stage when the colour preconfinement mechanism has already brought about 
the formation of {\em small} colour singlet clusters. These distinctive features
of hadronisation in HIC with respect to hadronisation in EC imply a difference of
relative strange quark production which is, hopefully (as we have assumed in (2)),
reflected into final hadrons. Moreover, the existence of a characteristic single
cluster volume in EC independent of colliding system and centre-of-mass energy 
is argued to be the main responsible for the constancy of $\lambda_S$ because of the
strange quark suppression entailed by the colour singlet constraint over 
a small spacial region (canonical colour suppression). The assumption (3) is 
certainly the strongest one, as it implies that hadronisation itself is assumed to
be a critical process and, thence, hadronisation temperature is to be identified 
with the critical QCD temperature; this is suggested by the observed constancy of 
fitted temperature for many collisions \cite{beca2} and by its value $T \simeq 160$ MeV
close to the calculated lattice value \cite{karsch}.   

As has been mentioned, effective lagrangian models are a useful tool to deal with 
quark degrees of freedom at finite temperature. In fact, none of them can account 
for colour confinement but many of them embody chiral symmetry breaking ($\chi$SB) 
and its restoration ($\chi$SR), which is expected to occur at the same critical point 
\cite{karsch}. The predictions of EM about temperature dependence of several 
physical quantities may strongly vary, nevertheless they also show striking common
features. Specifically, the phase diagram for $\chi$SR exhibits a tricritical point 
in the chiral limit $m_{q}\rightarrow 0$ separating second order from first order 
phase transitions \cite{bcd} and low-$\mu$ and high-$T$ region is second order.
Moreover, the expression for the number of quarks for the $i^{\rm th}$ flavour
generally stems from the one-loop term and, in the mean-field approximation of
four-fermion models, it can be derived from a free Dirac Hamiltonian with constituent 
quark masses replacing current masses, so that in the grand-canonical ensemble:

\begin{equation}\label{numgc}
\langle n_{i}\rangle = {N_{c}V \over \pi^{^{2}}} \int_{_{0}}^{^{\Lambda}} 
\d p \; \frac{p^2}{\exp [ \sqrt{p^2+M_{i}^2}/T-\mu_{i}/T]+1}
\end{equation}
where $N_c=3$ and $\mu_i$ are the relevant chemical potentials. In Eq.~(\ref{numgc}) 
$\Lambda$ is an UV cutoff which is needed in EM with four-fermion interactions as 
these models are not renormalisable (630 MeV in ref.~\cite{njl}). In {\it ladder}-QCD 
models essentially the same expressions holds below $\Lambda$ whereas at higher momenta 
the one-loop Hamiltonian has to be modified. However, this modification usually implies 
a minor change of the ratio s/u since the largest contribution to $\langle n_{i}\rangle$ 
comes from the integration region $p < \Lambda$.
Finally, commonly to most EM, the u and d quark constituent masses $M_{u,d}$ steeply 
decrease to a value close to the current mass within a small 
interval of temperature (which is fairly identified with the critical region) whereas 
the strange quark constituent mass $M_s$ decreases much more slowly and in fact, in 
the critical region, it has a value still much higher than current mass value.

In order to obtain quantitative predictions for EC within a given effective model
the canonical expressions of quark numbers are needed, hence we have to implement 
exact colour and flavour conservation over finite volumes. This task can be 
accomplished by means of well known methods based on group theory \cite{exact}. 
In our case the involved symmetry group is $G={\rm SU}(3)_{c} \times {\rm U}(1)_{u}
\times {\rm U}(1)_{d} \times {\rm U}(1)_{s}$ and physical states to be counted in
the partition function of a single cluster should be projected onto the irreducible
invariant 1-dimensional subspace associated with the colour singlet representation 
with given initial flavour numbers. The overall number of quarks for a given flavour
$i$ can be obtained by taking the derivative of the overall partition function (i.e.
the partition function of the multi-cluster system) with respect to fictitious 
fugacities $\lambda_i$:

\begin{equation}\label{ncan}
 \langle n_{i}\rangle = \frac{\partial \log Z (\lambda_i)}{\partial \lambda_i}
 \Bigg|_{\lambda_i=1}
\end{equation}
where the overall partition function reads (the proof is given in ref.~\cite{noi}):

\begin{eqnarray}\label{zeta}
&& Z(\lambda_1,\lambda_2,\lambda_3) = \left[ \prod_{i=1}^3 \int_{-\pi}^{\pi}
{\d \phi_{i} \over 2\pi} \exp [\i N_i \phi_i] \right] \Bigg\{ 
 \int \d \mu(\theta_1,\theta_2) \; \exp \Bigg[ \sum_{i=1}^3 {2 V_c \over (2 \pi)^3} 
 \nonumber \\ 
&& \times \sum_{n=1}^\infty {(-1)^{n+1}\over n} \, \chi_{1,0}(n\theta_1,n\theta_{2}) 
\, \e^{\i n \phi_i} \, \lambda_i^n \int_{_{0}}^{^{\Lambda}} \d^3 {\rm p} \; 
\exp [-\sqrt{p^2+M_i^2}/T] + {\rm c.c.} \Bigg] \Bigg\}^{V/V_c} 
\end{eqnarray}
where $\d \mu$ is the normalised SU(3)$_c$ group measure \cite{exact} and $\chi_{1,0}$ is 
the character of the fundamental SU(3)$_c$ representation. Two different volumes 
show up in Eq.~(\ref{zeta}) owing to the fact that colour singlet 
constraint applies to each single cluster (with volume $V_c$) whereas the flavour 
constraint (equal to that of initial state) applies to the system of clusters 
overall; indeed $V$ is meant to be the sum of all clusters proper volumes.    

\section{Analysis and Results}

In our numerical analysis we essentially determine $\lambda_S$ on the basis 
of Eqs.~(\ref{ncan}), (\ref{zeta}). Generally speaking, $\lambda_S$ depends 
on the value of constituent quark masses at a fixed temperature $T$, on the 
cutoff $\Lambda$ and on the two volumes $V$ and $V_c$. According to statement (3) 
in previous section, the temperature $T$ has been set equal to the value fitted
within SHM, which is about 160 MeV in EC and in Pb--Pb collisions \cite{becahion}. 

As far as HIC are concerned, $\lambda_S$ calculation can be performed in the 
grand-canonical ensemble on the basis of Eq.~(\ref{numgc}) because of the very 
large involved overall volumes $V$. Colour canonical suppression (see below) sets 
in only at very small $V_c$ volumes, which are excluded because of the assumption 
of colour deconfinement. Thereby, since $\lambda_S$ is a ratio of particle numbers,
the dependence on volumes vanishes and, by setting $\Lambda\rightarrow\infty$ in 
Eq.~(\ref{numgc}), it is found that one needs $M_s \sim 500$ MeV and $M_u < 100$ MeV 
to match the fitted $\lambda_S$ value in Pb--Pb ($\simeq 0.4$) at $T=160$ MeV. 
This means that an eligible effective model should have a critical region 
(at very low baryon chemical potential) around 160 MeV with constituent u and d 
masses essentially dropped from their value at $T=0$. These features can be 
precisely found in one version (case 2) of the NJL model in ref.~\cite{njl} 
which has thus been used to make a quantitative comparison
with the data. In particular, this model has an UV cutoff $\Lambda = 630$ MeV,
$M_{u,d}(T=160~{\rm MeV}, \mu=0) = 64$ MeV and $M_{s}(T=160~{\rm MeV}, \mu=0) 
= 449$ MeV. The agreement between calculated and fitted $\lambda_S$ values in Pb--Pb 
is very good as it can be seen in Table~1. 
\begin{table}[htb]
\begin{center}
\begin{tabular}{ccccc}
\hline
 Collision      &  $T$ (MeV)      & $\mu_q$ (MeV)   & $\lambda_S$       & $\lambda_S$ (calc.)  \\
\hline
  Pb--Pb SPS    & $158.1 \pm 3.2$ & $79.3  \pm 4.3$ & $0.447 \pm 0.025$ & 0.455 \\  
  Au--Au RHIC   & $165 \pm 7$     & $13.7 \pm 1.7$  &                   & 0.335 \\
\hline
\end{tabular}
\caption{Comparison between $\lambda_S$ fitted with SHM \cite{becahion} and the
calculated $\lambda_S$ in high energy heavy ion collisions by using the central fitted 
values of $T$ and $\mu_q \simeq\mu_B/3$ \cite{becahion, florkowski}}
\label{hions}
\end{center}
\end{table}

The effect of volume finitiness on $\lambda_S$ ({\em canonical suppression}) is
mainly relevant to EC and has been accurately studied by enforcing colour and 
flavour constraints either separately or simultaneously. Generally speaking, the 
requirement of an exact conservation of some global quantity (such as colour or 
charge) suppresses heavier particles more than lighter ones with 
respect to the grand-canonical limit because, at finite $T$, less energy can be 
spent to compensate the unbalance created by one particle generation. Indeed, for 
a given volume $V=V_c$, the canonical suppression of s quarks with respect to u, d 
quarks is expected, and has been found, 
to be predominantly determined by the net zero strangeness constraint rather than 
by colour singlet constraint as colour can be compensated by the generation of 
two light u, d quarks instead of one heavier $\bar {\rm s}$ quark. 
Unlike the number of quarks, the masses of constituent quarks have been determined 
by minimising the free energy in the grand-canonical limit, i.e. neglecting the 
effect of finite volume, at $T=160$ MeV and $\mu_i = 0$.   
\begin{figure}
\begin{center}
  \resizebox{20pc}{!}{\includegraphics{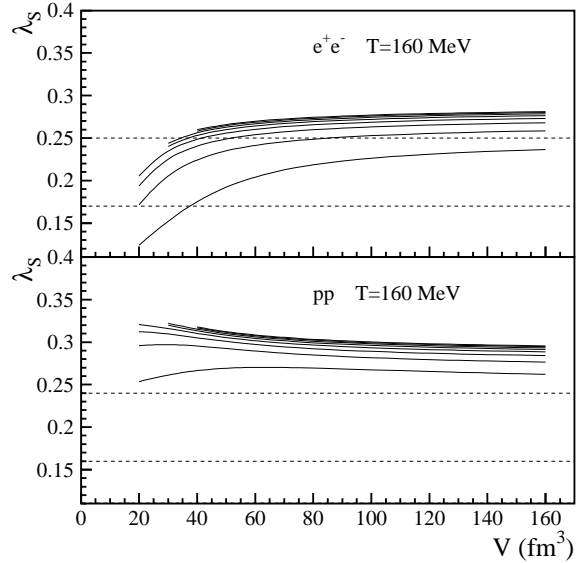}}
  \caption{Calculated $\lambda_S$ in \ee and pp collisions at $T$=160 MeV within
  the NJL model \cite{njl} as a function of the total volume $V$ for single 
  cluster volume $V_c$ varying from 5 fm$^3$ (lowest curves) to 40 fm$^3$ in 
  steps of 5 fm$^3$. The horizontal bands are the ranges of fitted
  $\lambda_S$ values in the SHM (see Fig.~1).}
\end{center}  
\end{figure}

The effect of combined colour and flavour conservation over different volumes on 
$\lambda_S$ is shown in Fig.~2 as a function of $V$ and $V_c$ for colliding systems 
such as \ee (flavour neutral) or pp, along with conservatively estimated ranges of 
$\lambda_S$ determined with the multiplicity fits within the SHM (see Fig.~1). 
The colour canonical suppression of $\lambda_S$ clearly shows up for single-cluster 
volumes below $\approx 15$ fm$^3$. Apparently, a $V_c$ between 5 and 10 fm$^3$ can 
account for the observed constant $\lambda_S$ value in \ee collisions whereas the 
predicted value in pp is too high. In fact, the relative strangeness enhancement 
due to the presence of initial u and d quarks in the colliding protons which 
inhibits the creation of \uu, \dd pairs, seems to prevail over the relative 
strangeness suppression entailed by colour and $S=0$ constraint. 

\section{Conclusions}

We have studied strangeness production within a statistical model of hadronisation
by using effective models with constituent quarks. The basic idea is that 
full statistical equilibrium is achieved at the level of quark degrees of freedom
in both elementary and heavy ion collisions. Besides the effect of different
initial light flavour content and density, the smaller relative strangeness
production in EC with respect to HIC is supposed to be related to the smaller
sytem size and to colour confinement over small distances. A quantitative study
in this regard gives a satisfactory agreement with the data in heavy ion and \ee 
collisions but a significant disagreement in pp, which might be cured by 
taking more complex assumptions about quantum numbers distribution among the 
hadronising clusters.  

 

\begin{thebibliography}{99}

\bibitem{beca}
  F. Becattini, Z. Phys. {\bf C69} (1996) 485;      
  F. Becattini and U. Heinz, Z. Phys. {\bf C76} (1997) 269.
\bibitem{beca2} F. Becattini, L. Bellucci, G. Passaleva, 
  Nucl. Phys. Proc. Suppl. {\bf 92} (2001) 137;  
  F. Becattini, G. Passaleva, in preparation.     
\bibitem{becahion} F. Becattini, M. Gazdzicki, J. Sollfrank, 
  Eur. Phys. J. {\bf C5} (1998) 143; F. Becattini et al., Phys. Rev.
  {\bf C64} (2001) 024901.
\bibitem{njlor}
  Y. Nambu, G. Jona-Lasinio, Phys. Rev. {\bf 122} (1961) 345;
  {\bf 124} (1961) 246.
\bibitem{njl}
  T. Hatsuda and T. Kunihiro, Phys. Rep. {\bf 247} (1994) 221.
\bibitem{bcd}
  A.Barducci, R.Casalbuoni, S.De Curtis, R.Gatto and G.Pettini,
  Phys. Rev. {\bf D41}, (1990) 1610; 
  A. Barducci, R. Casalbuoni, R. Gatto and G. Pettini, Phys. Rev. 
  {\bf D49}, (1994) 426 and references therein.
\bibitem{karsch} F. Karsch, this conference.  
\bibitem{exact}
     K. Redlich and L. Turko, Z. Phys. {\bf C5}, (1980) 201;
     M.I. Gorenstein et al., Phys. Lett. {\bf B123} (1983) 437; 
     G. Auberson et al., J. Math. Phys. {\bf 27} (6) (1986) 1658.
\bibitem{noi} 
     F. Becattini and G. Pettini, in preparation.
\bibitem{florkowski} 
     W. Florkowski, W. Broniowski and M. Michalec, nucl-th/0106009.
\end{thebibliography}
\end{document}